\documentstyle[prl,aps,psfig,floats,twocolumn]{revtex}
\begin{document} 
\draft 
\twocolumn[\hsize\textwidth\columnwidth\hsize\csname
@twocolumnfalse\endcsname
 
\title{Dynamics of Fractures in Quenched Disordered Media.} 
\author{Guido Caldarelli$^{1,2}$, Raffaele Cafiero$^{3,4}$  
and Andrea Gabrielli$^{4,5}$} 
\address{$^1$Theory of Condensed Matter, Cavendish Laboratory Madingley Road
CB3 0HE Cambridge UK
} 
\address{ 
$^2$Department of Theoretical Physics, The University, M13 9PL  
Manchester, U.K.} 
\address{ 
$^3$Max-Planck Institute for Physics of Complex Systems, 
N\"ohnitzer strasse 38, 01187 Dresden (Germany)} 
\address{ 
$^4$INFM-unit\`a di Roma I, Universit\`a di Roma "La Sapienza", 
P.le Aldo Moro 2,I-00185 Roma, Italy,} 
\address{ 
$^5$Dipartimento di Fisica, Universit\'a di Roma "Tor Vergata", 
via della Ricerca Scientifica 1, 00133 Roma, Italy.} 
\maketitle 
\date{\today} 
\maketitle 
 
\begin{abstract} 
We introduce  a model for fractures in quenched disordered media. 
This model has a deterministic extremal dynamics, driven by the energy  
function 
of a network of springs (Born Hamiltonian). The breakdown is the result  
of the  
cooperation between the external field and the quenched disorder.  
This model can be considered as describing the low temperature limit for 
crack propagation in solids. To describe the memory effects in this dynamics, 
and then to 
study the resistance properties of the system we realized some numerical  
simulations of the model. The model exhibits interesting geometric and 
dynamical properties, with a {\em strong reduction} of the fractal dimension 
of the clusters and of their backbone, with respect to the case in which 
thermal fluctuations dominate. This result can be explained by a recently 
introduced theoretical tool as a {\em screening enhancement} due to memory 
effects induced by the quenched disorder.
\end{abstract} 
 
]
\narrowtext

\section{Introduction} 
In the last years, many models have been proposed to describe  
the formation of cracks in different kind of materials  
\cite{HerrRoux,vicsek,herr,zapp}. They are inspired by the study of  
non equilibrium fractal growth processes, like the Dielectric  
Breakdown Model (DBM) \cite{dbm} and Diffusion Limited  
Aggregation (DLA) \cite{dla}. 
These models are based on two different mechanisms for the fracture 
growth:  
1) A {\em stochastic} effect due to the thermal fluctuations in the medium,  
driven by an external field.  
2) A {\em deterministic} dynamics, when the main source of randomness 
is the quenched disorder of the medium (like in the Invasion Percolation (IP) 
model \cite{wilk}. ), being the thermal fluctuations negligible  
(i.e. low temperature limit). 
 
A very interesting model, belonging to the first class, is the Born Model  
(BM) \cite{bm,bmguido}, where minimization of the elastic energy is used to  
compute the driving field.  
This model for fracture propagation is the analogous of the DBM 
for Laplacian growth. It describes the fractures  
propagation as a stochastic process, where the probabilistic  
mechanism represents growth instabilities, like for  
example density fluctuations in a gas. In this approach the quenched disorder 
of the medium is negligible. 
 
In this paper we want to consider the limit of low temperature for the  
BM, where the driving field cooperates  
with quenched disorder to produce the breakdown patterns. 
The quenched disorder can be thought to represent the effect, at a mesoscopic scale, of defects of the breaking layer. 
In this version of the model, at each time step the growing bond is selected  
deterministically, i.e. the bond with the smallest ratio between the local  
driving field and the quenched disorder grows.  
The system is a $2d$ triangular lattice of springs. 
We apply to two parallel boundaries of the system an uniaxial tension  
and fixed boundary condition to the others. 
In this way we obtain directed, crack patterns, orthogonal to the applied  
stress.  
The system has two independent length scales, the height $h$  
and the width $L$. This allow us to evaluate more clearly the various  
scaling regimes of these structures.  
Usually, with this boundary conditions, two different phases, called  
respectively {\em scaling regime} and 
{\em steady state}, are present \cite{evertsz}.  
 
We will see that in quenched version of BM (QBM), because of the very  
strong screening on the growth process, the first regime (where many  
different branches compete during the growth) is nearly absent.  
In fact, most of the growth dynamics develops during the steady state, where 
only one branch survives and the cluster is statistically self-similar  
with well defined fractal properties.  
Together with the fractal dimension, the backbone and chemical  
distance exponents characterize completely the fractures produced by  
our model.  
 
We compare our results with those obtained previously for the stochastic  
version of the BM \cite{bmguido} showing that the fractal dimension  
of the clusters and of the backbone,  
in similar conditions, are {\em strongly reduced}. 
We explain this as a consequence of the absence of thermal fluctuations;  
the quenched disorder produces {\em memory effects} giving screening effects  
similar to those of IP \cite{rtslung}. 
On the other hand, in the same conditions, the screening  
effect produced by the modulation of the  
driving field gives a fractal dimensions much smaller than IP,  
which is the limit of the model in which there is no external driving field  
or it is constant.   
The same qualitative result has been recently found, both numerically  
and analytically, for the quenched version of the DBM model  
\cite{qdbmrts}.  
Therefore, the numerical results found here for the QBM 
support the feeling that this screening enhancement is  
a general property of all deterministic models with quenched disorder and 
a driving field.  
  
The distribution of the quenched disorder related to grown (broken) sites  
(acceptance profile) is also studied for different values of the model  
parameters. 
This distribution reaches, during the evolution, an asymptotic shape.  
In the case of Invasion Percolation, 
the asymptotic distribution is a step function, with the discontinuity at the  
value of quenched disorder coincident with the critical point of classic  
percolation \cite{wilk,rts}. This step-like shape of the acceptation 
profile indicates that the dynamics develops avalanches with a scale 
invariant distribution \cite{gap,rtslung}. 
  
In our model, the presence of the stress field, however, does not allow  
the presence of a critical threshold in this asymptotic distribution of  
quenched disorder.  
As a consequence, the dynamics does not develop scale invariant  
avalanches, 
but these avalanches have a typical size \cite{rtslung,qdbmrts}. 
A very important universality relation can be written, which explains, 
in terms of the dependance of the fractal properties on the 
parameters of the model, the cooperation between the driving field and the  
quenched disorder in developing such a fractal structure.  
 
The paper is organized as follows. In section II the model is introduced  
and compared with the corresponding stochastic model. The details for  
the realization of simulations are specified. In section III we describe  
our numerical results for the fractal properties of the clusters, the  
backbone and the chemical distance, for the roughness exponent  
of the chemical  
distance, and for the statistical effective  
distribution of quenched variables on the growth interface. In section  
IV the universality relation is demonstrated and a theoretical tool is  
introduced to explain the two screening effects in producing fractal  
properties.   
Finally in section V we discuss the results and draw some conclusions. 
 
\section{The model} 
 
The BM  describes the medium as a discrete set of springs. 
The equilibrium state is obtained by imposing minimization of the energy while 
dynamics of fracture is given by assigning a rule of growth. 
For what concerns the equilibrium state,
we imagine to break only one spring a time to model a system of slowly 
developing fractures, and
after every breakdown a new equilibrium state is computed. 
We describe the energy of the system 
by means of the same potential energy. 
This elastic potential energy consists of two different terms, describing, 
respectively, a central force and a non-central force contribution: 
\begin{eqnarray} 
V&=&\frac{1}{2} \sum_{i,j} V_{i,j} = \nonumber \\  
 &=&\frac{1}{2} \sum_{i,j} (\alpha - \beta) [(\vec{u}_i- 
\vec{u}_j) \cdot \hat{r}_{i,j} ]^2 
 +  \beta [\vec{u}_i-\vec{u}_j]^2, 
\label{poten} 
\end{eqnarray} 
where $\vec{u}_i$ is the displacement vector for site $i$,  
$\hat{r}_{i,j}$ is the  
unity vector between $i$ and $j$, $\alpha$ and $\beta$ are force  
constants, and the sum is over the nearest neighbor sites  
connected by a non-broken spring. 
 
For any equilibrium state must result $\vec{\nabla}_{\{\vec{u}_i\}}  
V(\vec{u}_i,\vec{u}_j) = 0, \forall i$, where $j$ n.n. of $i$. 
This condition yields to a series of equation for the $\vec{u}_i$'s that 
can be solved imposing the boundary condition. 
The initial boundary condition is a uniform dilation on the left and 
right side of the sample. 
This boundary condition change, taking into account all the springs broken  
during the evolution of the crack.  
 
For what concerns the rule of growth we have chosen a deterministic rule
selecting the bonds to break according to its ``generalized elongation''
$V^{1/2}_{i,j}$. The particular rule explained in the following equations 
has been inspired by the analogies with the DBM model. 
We think $V_{i,j}^{1/2}$ as  
a field acting on the spring between sites $i$ and $j$.
Since the system is characterized by the presence of random quenched  
defects (represented as a quenched random noise),
we assign to each spring a random number $x_{i,j}$ extracted by the 
probability density: 
\begin{equation} 
p_0 (x_{i,j})=\frac{a}{x_{sup}^a} x_{i,j}^{a-1}, 
\label{quench} 
\end{equation} 
where the parameter $a$ ($a \in [0,\infty)$) 
modulates the importance of the disorder in the  
mechanical properties of the system,  
and the variables are defined in the range $[0,x_{sup}]$.  
From eq. \ref{quench} one can derive the mean value $\langle{x}\rangle$ 
of the disorder assigned to the bonds:
\begin{equation}
\langle{x}\rangle=\int_0^{x_{sup}}dx x p_0(x)=\frac{a}{a+1}x_{sup}.
\label{xmed}
\end{equation}
A ``fragile" 
material correspond to small values of $a$ ($\langle{x}\rangle \simeq 0$), 
while a ``rigid" one corresponds to big values of $a$ 
($\langle{x}\rangle \simeq x_{sup}$). 
Then we define a set of dynamical variables: 
\begin{equation} 
y_{i,j}(t)=A_{i,j}(t) x_{i,j} 
\end{equation} 
where $A_{i,j}(t)=\frac{1}{V_{i,j}(t)^{1/2}}$. 
Then, at each time step, the spring with the smallest value $y_{i,j}(t)$  
on the growth interface of the fracture is broken.

\begin{figure} 
\centerline{\psfig{file=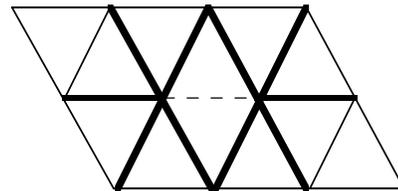,height=2.5cm}} 
\caption{Connectivity condition. Broken bonds are  
indicated by dashed lines, 
while interface bonds are indicated by thick lines.} 
\label{fig1} 
\end{figure}

It is customary to modulate the influence of the stress  
field on the growth dynamics with the introduction of a parameter
$\eta$. In this case the formula for $A_{i,j}$ becomes 
$A_{i,j}(t)=\frac{1}{V_{i,j}(t)^{\eta/2}}$.
When $\eta=0$ there is no field and the model 
has the same dynamics as invasion percolation \cite{qdbmrts}, while  
$\eta=\infty$ correspond to an infinite strength of the field and  
the cluster is an one-dimensional straight line.  
In fact, the disorder effects are negligible when compared to an infinite 
field.  
Some details have to be specified since in principle their 
variation could affect the fractal properties \cite{bm,cfm,meak}.  
Due to the vectorial nature of eq.(\ref{poten}) a triangular lattice 
is more appropriate to model the medium. In fact, for a  
squared lattice when $\beta=0$, the system 
behaves as a set of independent planes without connection between each other. 
For this reason we will follow \cite{bmguido,guido2}, by considering 
only triangular lattices. 
Furthermore, the growth interface at any time is given by the set of non  
broken bonds nearest neighbors to the cluster ones (i.e. the perimeter of  
the fracture cluster). This corresponds to the implementation of a 
connectivity condition (FIG. \ref{fig1}). 
 
\section{Numerical simulations} 
We performed several realizations on systems of size $L \times L$  
($L=64,128$), in triangular geometry (with periodic boundary conditions 
on top and bottom sides). 
 
\begin{figure} 
\centerline{\psfig{file=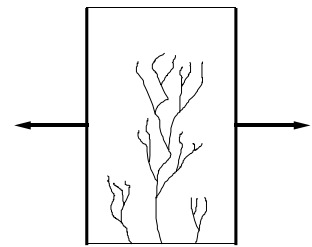,height=4.0cm}} 
\caption{Uniaxial tension applied to the lattice.} 
\label{fig2} 
\end{figure}

\begin{figure} 
\centerline{\psfig{file=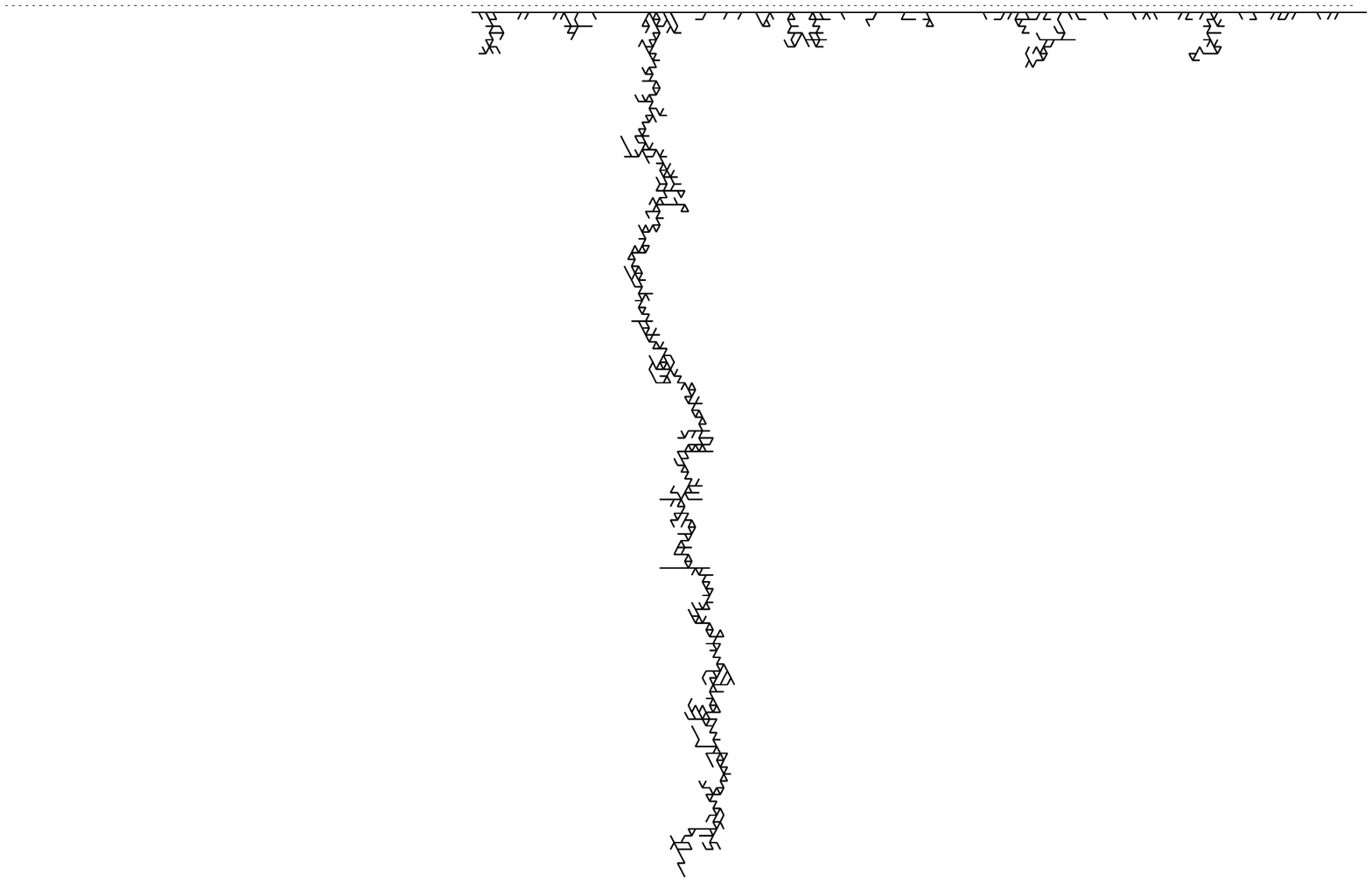,width=7.0cm,angle=-180}} 
\caption{A crack pattern generated by our model, with parameters $\eta=1.0, a=1.0, \alpha=1.0,\beta=0.5$.} 
\label{fig3} 
\end{figure}

A quenched variable $x_{i,j}$ is assigned to each bond ${i,j}$. 
The $x_{i,j}$'s follow the distribution function of  
eq.(\ref{quench}) with $x_{sup}=0.5$. In fact, a constant strain  
equal to $0.5$ lattice units is applied in the horizontal direction 
(see FIG.\ref{fig2}).  
At each time step, the stress field over the interface bonds is  
computed. Then the bond with the smallest value $y_{i,j}$ is broken, 
the new stress fields are computed and new bonds are added to the perimeter. 
By iterating this dynamics, one obtains structures like that shown 
in FIG. \ref{fig3}.

The most important quantity characterizing the structures  
generated by the model is the fractal dimension $D_f$.  
For each value of $L$, we have generated a set of $20$ realizations and  
 computed the fractal dimension by the box-counting method.  
This has been done for different values of $\beta/\alpha$. 
In fact, in the equilibrium condition $\vec{\nabla}_{\{\vec{u}_i\}} 
V(\vec{u}_i) = 0$ we deal only with the ratio $\beta/\alpha$. 
For this reason, we decided to vary $\beta$ and to keep $\alpha=1$. 
Furthermore, we performed the same analysis also by varying $\eta$  
(the parameter modulating the effects of the field) and $a$  
(the parameter modulating the disorder).  
Our results are shown in Tab. \ref{tabdf} and Tab. \ref{tabdf1}, 
 \ref{tabdf2}.  
  
The model shows a continuous dependence of the fractal 
dimension on the parameter $\beta$, as found in \cite{bmguido} for the 
stochastic BM. The dependance of the fractal dimension on the  
parameters $\eta$ and $a$ is, instead, new and interesting.  
In fact, from our simulations we see that the fractal dimension  
actually {\em depends only on the product $a\eta$}. 
This introduces a precise relationship between the indices 
describing the properties of the medium and the properties 
of the stress field. 
 
If we compare this result with the case of invasion 
percolation, where the field is absent and the fractal dimension does not 
depend on the value of $a$ \cite{wilk,rtslung}, we see that the  
introduction of the stress field breaks the symmetry with respect to $a$,  
leading to less universal ``critical'' properties.  
In the following section we present an analytical demonstration of 
this universal property. 
 
As it can be noted, for fixed $a$, the fractal dimension of the fracture  
cluster, for any value of $\eta > 0$, is less than the IP one. 
This is due to the fact that for $\eta > 0$, there is a  
screening effect related to the physical stress field, in addition to  
the screening related to the memory effects of the quenched disorder. 
Moreover, our numerical results show clearly that in a ``fragile" material 
(small $a$, $\langle{x}\rangle \simeq 0$)
the fractures have a big fractal dimension, while in a ``rigid" material 
 (big $a$, $\langle{x}\rangle \simeq x_{sup}$) the fractures 
tend to be straight lines with fractal dimension near to $1$. 
This result sounds qualitatively reasonable from an experimental point 
of view. A ``fragile" material could correspond to a material with many 
impurities which lower its resistance to rupture, allowing many 
bonds to be broken. A rigid material could represent a material 
without impurities at zero temperature, in which fractures are straight 
lines along the direction of maximum stress. 
However, a quantitative comparison with experiments is still  
not accessible, since it needs a clear connection 
between what we call quenched disorder in our model, which we believe 
to give a description of the system at a mesoscopic scale, and 
the microscopic disorder in real systems. 
 
In the next section we will propose 
an analytical explanation for our numerical findings. The same 
qualitative results have been recently found  
for a similar model, the quenched dielectric breakdown model (QDBM)  
\cite{qdbmrts}.  
This suggests that the dependance on the product of $a$ and $\eta$  
of the dynamics and geometry of the patterns plus a {\em screening enhancement} 
effect are general properties of all the deterministic models with  
driving field in presence of quenched disorder. 
 
\begin{figure} 
\centerline{\psfig{file=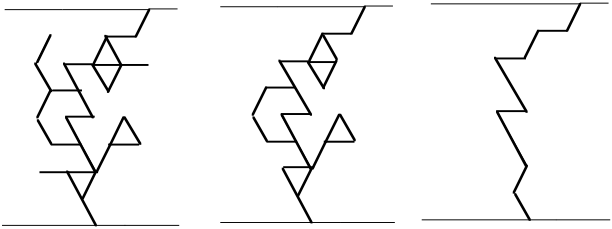,height=3.0cm,angle=90}} 
\caption{A schematic representation of an aggregate (left), 
of its backbone (middle)  
and of its chemical distance (right).} 
\label{fig4} 
\end{figure}

A further characterization of the topological and connectivity properties of 
the aggregates is given by the exponents ruling the scaling of two 
subsets of the clusters: the chemical distance and the backbone. 
The chemical distance is the shortest path between the two ends of the  
aggregate and shows interesting self-affine properties.  
The backbone is obtained cutting from the cluster all the  
tips and the dangling loops connected to the chemical distance   
(see FIG. \ref{fig4}).  
This part of the cluster influences the macroscopic  
transport properties of the medium, while the chemical distance gives 
the shape of the line separating the system in two parts 
after the breakdown. It is worth to point out that in models with 
a scalar external field, like DBM \cite{dbm} and QDBM \cite{qdbmrts}, 
no closed loops are allowed, and the backbone trivially coincides 
with the chemical distance \cite{bmguido}.
 
\begin{figure} 
\centerline{\psfig{file=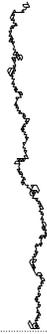,width=7.0cm,angle=-180}} 
\caption{The backbone of the cluster shown in Fig.3.} 
\label{fig5} 
\end{figure}

\begin{figure} 
\centerline{\psfig{file=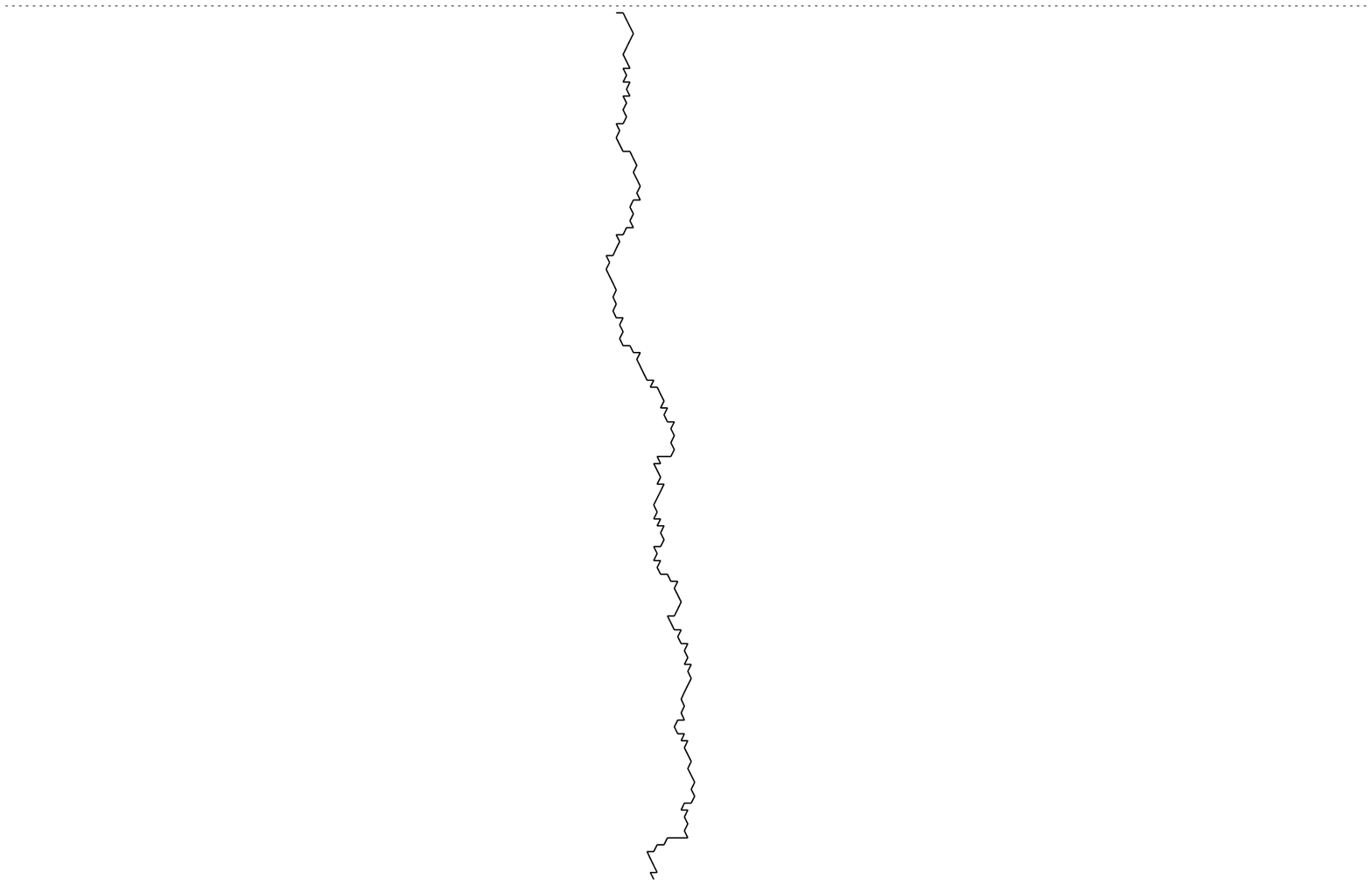,width=7.0cm,angle=-180}} 
\caption{The chemical distance of the cluster of Fig.3.} 
\label{fig6} 
\end{figure}

To determine the backbone and the chemical distance 
we followed a method used in \cite{bmguido}, based on the topological  
properties of the clusters. 
As an example, in FIGG. \ref{fig5}, \ref{fig6} we show, 
respectively, the backbone and the chemical distance 
we obtain for the cluster of FIG. \ref{fig3}. In Tab. \ref{tabdf3} 
we collect the results of a 
box-counting analysis. The backbone fractal dimension is  
significantly larger than one, although 
quite smaller than for the stochastic Born model \cite{bmguido}. 
Instead, the fractal dimension of the 
chemical distance is $1$ for both $L=64$ and $L=128$.  
 
\begin{figure} 
\centerline{\psfig{file=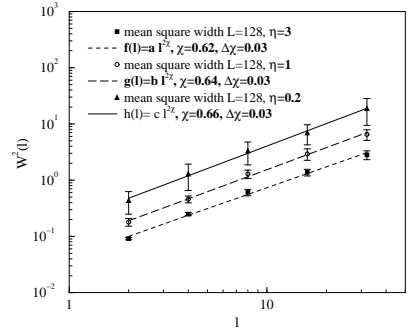,height=5.0cm,angle=0}} 
\caption{ 
Scaling behaviour of the square mean horizontal  
width $W^2(l)$ of the chemical distance for clusters of size $L=128$ and different values of $\eta$.} 
 \label{fig7} 
 \end{figure}

We then studied the roughness exponent $\chi$ of the chemical distance,  
which gives the scaling of the mean squared lateral width $W(l)$  
of a self-affine path with respect to its length $l$.  
In our case, the path 
develops vertically, and its length is measured along the vertical direction, 
while the width is computed along the horizontal direction.  
Then, if $i(j)$ is the $x$-coordinate of a point on a  
chemical distance whose $y$-coordinate is j, $W(l)$ is defined as:  
\begin{equation} 
W(l)=\left[ \langle{ 
\sum_{j=j_0}^{j_0+l} (i(j)-\bar{i})^2 \rangle} \right]^{\frac{1}{2}} 
\label{rough} 
\end{equation} 
where $\langle{...}\rangle$ represents a mean over all portions of length $l$ 
of the chemical distance and over different realizations  
of quenched disorder, and  
$\bar{i}$ is the mean horizontal position.  
For a self-affine path $W(l)\sim l^{\chi}$ holds. 
In FIG. \ref{fig7} we show the scaling behaviour of $W^2(l)$ versus 
$l$, for clusters of size $L=128$ and for different values  
of the parameter $\eta$. A least square fit, reported 
in the figure, gives for values of $\eta$ ranging over an order of 
magnitude (from $0.2$ to $3$), values which seem quite stable 
within error bars. For $\eta=3$ we get $\chi=0.62\pm0.03$, 
for $\eta=1.0$ we get $\chi=0.64\pm0.03$ and for $\eta=0.2$ we get 
$\chi=0.66\pm0.03$. 

The independence of  $\chi$ from $\eta$ (i.e. from the product  
$a \eta$) indicates a very important universality property of the model. 
In particular the values  we find are quite similar to that of  
the directed percolation ($\chi=\frac{\nu_{\parallel}}
{\nu_{\vert}}=0.625...$ \cite{zhang}),  
so we can suppose that, for what concerns the  
chemical distance, the model is in the same universality class of  
 directed percolation. 
 
\begin{figure} 
\centerline{\psfig{file=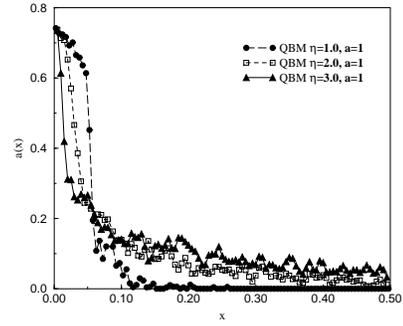,height=5.0cm}} 
\caption{Acceptation profiles for the variables  
$x_i$ associated to the grown bonds, for different values of $\eta$,  
after $t=500$ time steps.  
The absence of a threshold value for  
$a(x)$ implies a smooth growth (not by avalanches). Moreover, one can note  
a dependance on the value of $\eta$, suggesting that in the limit $\eta=0$  
we recover the IP dynamics with a threshold.} 
 \label{fig8} 
 \end{figure}

The evolution of the dynamics towards an asymptotic stationary state 
 is characterized by the acceptance profile $a(x)$. It   
gives the rate of acceptation (selection by the growth process) 
for the values of the quenched variables associated to the grown bonds,  
in the  
range $[x,x+dx]$. We have performed 
 a set of $10$ realizations of size $512\times512$, 
for $\eta=1.0,2.0,3.0$ and $a=1,\beta=0.5$. This allowed us to follow the  
time evolution of $a(x)$ up to $t=500$, where the asymptotic state 
is reached. In FIG. \ref{fig8} we show the final shape  
of $a(x)$, for the different values of $\eta$. The presence of  
the stress field prevents the dynamics from 
developing scale invariant avalanches and this is reflected in the absence of 
 a critical threshold in $a(x)$ \cite{rtslung}.  
However, as the value of $\eta$ becomes 
smaller, and the role of the stress field becomes less relevant, one can see 
 that $a(x)$ seems to develop a discontinuity. What we expect is that in  
the limit $\eta=0$ one recovers the IP dynamics.  
 Also in this case we obtain  
results that agree with those obtained for the QDBM model \cite{qdbmrts}. 
In fact, a common picture, that we describe in a longer paper,
 can be used to study both  models \cite{qdbmlung}. 
 
\section{Theoretical results} 
 
In this section we present some theoretical results concerning the universality properties of the model and the explanation of the screening effects which 
give rise to such fractal breakdown patterns.  
 
To demonstrate that all the dynamical and geometrical properties of the  
model depends only on the product $a \eta$ instead on the two parameters  
separately, consider the generic variable 
\begin{equation} 
y_{i,j}(t)= \frac{x_{i,j}}{V_{i,j}(t)^{\eta/2}} 
\label{defy} 
\end{equation} 
where $x_{i,j}$ is extracted from the density function 
 
\begin{equation} 
p_0 (x_{i,j})=\frac{a}{x_{sup}^a} x_{i,j}^{a-1} 
\label{dens1} 
\end{equation} 
 
Now perform the following transformation of the quenched variables 
\begin{equation} 
z_{i,j}=[x_{i,j}]^a 
\label{defz} 
\end{equation} 
 
Introducing eq.(\ref{defz}) in eq.(\ref{dens1}) one can note that 
the variables  
$ z_{i,j}$ are uniformly distributed between $0$ and $z_{sup}= x_{sup}^a$. 
So the density function of the new variables does not depend on the parameter 
$a$ apart from the value of  $z_{sup}$ (unique for all the bonds) value. 
Introducing eq.(\ref{defz}) in eq.(\ref{defy}) we obtain 
\begin{equation} 
y_{i,j}(t)= \left[\frac{z_{i,j}}{V_{i,j}(t)^{(a \eta)/2}}\right]^ {1/a} 
\label{defy2} 
\end{equation} 
Finally, as the dynamics is extremal it does not change if instead of the  
variables $y_{i,j}(t)$ we consider as  bond variables  
\begin{equation} 
u_{i,j}(t)= [ y_{i,j}(t)]^a =  \frac{z_{i,j}}{V_{i,j}(t)^{(a \eta)/2}} 
\label{defu} 
\end{equation} 
 
 This equation shows that all the dynamical and geometrical   
properties of the model depends only on the product $a \eta$ as   
asserted before.  
  
Now we switch to the explanation of how the two screening 
effects (the geometrical one  
and the probabilistic one) cooperate in the model during 
the dynamical evolution.
In doing so we will briefly introduce a generalized version of the 
Run Time Statistics (RTS), which was first introduced to study extremal 
dynamics in quenched disordered media, like IP (\cite{rts}, \cite{rtslung}). 
The RTS approach consists in a transformation of the deterministic quenched
extremal dynamics in a stochastic one through the introduction of
effective, time-dependent, density functions for the dynamical variables 
$ y_{i,j}$. The role of these densities is to store information on the 
past growth history of the system \cite{rtslung,rts}.

This approach allows us to have a  well defined growth probability
distribution at any time step and an updating rule of the effective density
functions after any elementary growth event.

\begin{figure} 
\centerline{\psfig{file=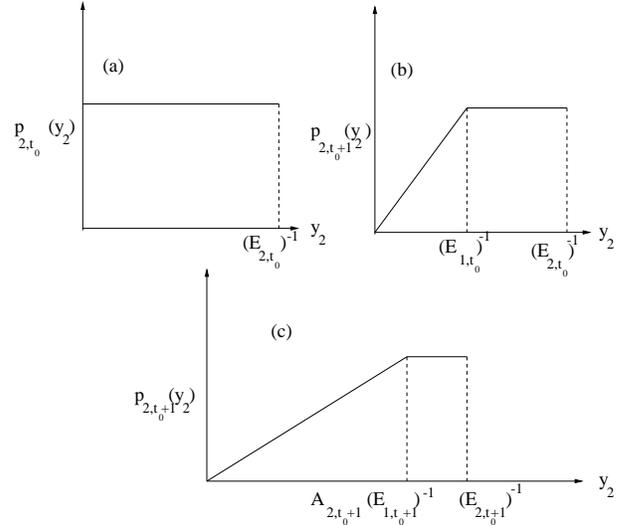,height=7.0cm,angle=-90}} 
\caption{(a) density function of bond $2$ at time $t_0$; (b) density function $p_{2,t_0+1}(y_2)$ of bond $2$ at time $t_0+1$, under the hyphotesis of constant stress field. This density concentrates more on high values. (c) If one adds the geometric screening of the stress field, the $p_{2,t_0+1}(y_2)$ concentrates even more on higher values, reflecting the cooperation of the two kinds of screening.} 
\label{fig9} 
\end{figure}

To explain simply the observed screening effects, suppose  to have
 only two bonds, 1 and 2,  
in the interface at time $t_0$ and consider $a=1$ and $\eta>0$.
 Let  the quenched variables of the two  
bonds be respectively $x_1$ and $x_2$, which are uniformly distributed  
between $0$ and $1$. Moreover let the related stress fields be  
respectively $E_{1,t_0} =V_{1}(t)^{\eta/2}$  
and  $E_{2,t_0}=V_{2}(t)^{ \eta/2}$ where for example, without loss 
of generality, $E_{1,t_0}\ge E_{2,t_0}$. 
Then we have  $y_{i,t_0}=x_i/E_{i,t_0}$ for $i=1,2$ 
uniformly distributed between $0$ and $1/E_{i,t_0}$ , i.e. 
\begin{equation} 
p_{i,t_0}(y)= E_{i,t_0}    \;\;\;   i=1,2 
\label{ro1} 
\end{equation} 
$p_{i,t_0}(y)$ represents the effective density of the bond $i$ at time $t_0$.
 We want now to calculate the probability $\mu_{1,t_0}$ that bond 
$1$ grows at $t_0 + 1$,  
i.e. that $y_{1,t_0}< y_{2,t_0}$. This probability is given by: 
\begin{equation} 
\mu_{1,t_0}\!\!=\!\!\!\int_{0}^{1/E_{1,t_0}}\!\!\!\!\!\!\!\!dy_1 
E_{1,t_0}\!\!\int_{y_1}^ {1/E_{2,t_0}}\!\!\!\!\!\!\!\! 
dy_2 E_{2,t_0}\!=\!1\!-\!\frac{E_{2,t_0}}{2 E_{1,t_0}}\! \ge \!\frac{1}{2} 
\label{mu1} 
\end{equation} 
Now, we suppose this is the real growth event and compute the new probability 
density  
$p_{2,t_0 +1}(y_2)$ of the variable $y_{2,t_0 +1}$ conditioned to that event. 
For sake of simplicity, we assume that the stress field  
of the bond $2$ keeps constant, i.e. 
$y_{2,t_0 +1}=y_{2,t_0}$. By applying the 
rules of conditional probability, we obtain: 
\begin{equation} 
p_{2,t_0 +1}(y_2)\!=\!p_{2,r_0}(y_2)\! \int_{0}^{y_2}\!\!\!dy_1 p_{1,t_0}(y_1)  
\theta(1/E_{1,t_0}\!-\!y_1) 
\label{ro2} 
\end{equation} 
where the function $\theta$ is the step function and indicates that 
the variable  
$y_1$ must be smaller than $1/E_{1,t_0}$. 
By introducing eq. \ref{ro1}  in eq. \ref{ro2} we obtain: 
\begin{equation} 
p_{2,t_0 +1}(y_2)= {\large\{} 
\begin{array} {ll}
\frac{E_{2,t_0}E_{1,t_0}}{\mu_{1,t_0}} y_2 \; ; 0 \le y_2 
\le \frac{1}{E_{1,t_0}}\\ 
\frac{E_{2,t_0}}{\mu_{1,t_0}} \; ; \frac{1}{E_{1,t_0}}\le y_2 
\le \frac{1}{E{2,t_0}} \\ 
\end{array} 
\label{ro3} 
\end{equation} 
This function is represented in FIG. \ref{fig9}-b and, 
the effective  probability density of  $y_2$ at time $t_0 +1$ 
is more concentrated towards the  
high values of the variable than at time $t_0$ at which was uniform. 
If we introduce this density in the eq. \ref{mu1}, we see that the 
growth probability of bond $1$ increases, and by normalization 
the growth probability of bond $2$ decreases.
This is what is called before "probabilistic screening effect", which is 
nothing else than a memory effect. 
In deriving the eq.(\ref{ro3}), we made the hypothesis that the stress field
of the bond $2$ remains constant. In general this is not the case, in fact when 
a bond $i$ grows at time $t$ the stress field of a bond $j$ remaining on the 
growth-interface decreases, i.e. $E_{j,t+1}\le E_{j,t}$.
This leads to the second kind of screening: the geometric one.
In fact, supposing $E_{2,t_0+1}\le E_{2,t_0}$, we obtain:
\begin{equation}
y_{2,t_0+1}=\frac{ E_{2,t_0}}{E_{2,t_0+1}} y_{2,t_0}=A_{2,t_0+1} y_{2,t_0}
\label{update}
\end{equation}
Note that, as $A_{2,t_0+1}\le 1$, we have $y_{2,t_0+1}\le y_{2,t_0}$.
From eqs.(\ref{ro3}) and (\ref{update}) we obtain the effective probability
density of the dynamical variable of the bond $2$:
\begin{equation}
p_{2,t_0 +1}(y_2)= {\large\{} \begin{array} {ll}
\frac{E_{2,t_0+1}E_{1,t_0}}{\mu_{1,t_0}} \frac{y_2}{A_{2,t_0+1}}   
\; ; 0 \le y_2 \le \frac{A_{2,t_0+1}}{E_{1,t_0}}\\
\frac{E_{2,t_0+1}}{\mu_{1,t_0}} \; ; \frac{A_{2,t_0+1}}
{E_{1,t_0}} \le y_2 \le \frac{1}{E{2,t_0+1}} \\
\end{array}
\label{ro4}
\end{equation}
where now we account for the decrease of the stress field in the variable 
$y_2$. Since, by definition, $A_{2,t_0+1}>1$, this function 
is more concentrated 
towards the high values of the variable than in the previous case (Fig. \ref{fig9}-c).
Thus, the growth probability of bond $2$ further decreases with respect to 
eq.(\ref{ro3}). 
This reflects the presence of both the screening effects, the geometric one 
due to the decreasing stress fields of "old" perimeter bonds, and the 
temporal one produced by memory effects.

In this way we have shown simply how the two kinds of screening effects
add the evolution of the dynamics leading to fractal
structure with small fractal dimensions. 

The generalization of these arguments to the case in which the interface 
is composed by many bonds with different initial probability densities,
leads to the formulation of the RTS 
for extremal dynamics with quenched disorder and a driving field, 
which has been introduced in \cite{qdbmrts} to study the problem of the 
electric discharge in disordered dielectric system. 

\section{Conclusions}
We have introduced and studied a model for the dynamics of fractures in a 
quenched random medium. The model exhibits interesting properties, 
for example a strong reduction of its fractal dimension with respect to 
the case of a dominating thermal noise (stochastic dynamics \cite{bmguido}), 
and a roughness exponent for the chemical distance whose value seems to be 
independent on the model parameters. Moreover, the fractal dimension 
of the fractures depends continuously on the material properties, represented 
in the model by a power law distribution of quenched disorder with 
a tunable parameter $a$. This is, from an experimental point of 
view, reasonable. In a situation in which thermal fluctuations 
of the lattice are irrelevant, the only source of noise, and 
thus of fractal properties, in the growth process is the 
intrinsic quenched disorder of the samples (impurities, vacancies etc.). 
Our model, however, is still too an idealized one to allow a 
quantitative comparison with experiments. 

It would be interesting to compute 
analytically the fractal dimension of this model by using the Fixed Scale 
Transformation (FST \cite{fst2}) approach, after having mapped its dynamics 
onto a stochastic one, with the RTS method. The mapping is in fact possible, 
but the approximations one has to use in the FST approach give rise to quite 
poor numerical results, as reported in a longer paper on this subject 
\cite{qdbmlung}. 
  
The model can be further generalized. A research direction we are presently 
following concerns the unification of the low temperature (deterministic, 
extremal dynamics with quenched disorder) and  high temperature 
(stochastic dynamics with thermal noise) regimes, by the introduction of a 
temperature-like parameter $T$ which can tune the transition between the two 
regimes \cite{qdbmgen}. This could represent a further step toward a 
more realistic description of fracture propagation in solids.
One interesting property of this generalized model is 
that in the high temperature limit, the level of approximation of the FST 
method becomes excellent \cite{fst2}. This could allow us to get a good 
estimation of the low temperature fractal dimension by an extrapolation 
of the values we get in the high temperature case.

\begin{table} 
\begin{centering} 
\protect \caption{Behavior of the fractal dimension of the QBM model 
for different values of $\beta$, with $a=1, \eta=1$, for sizes $L=64,128$.} 
\label{tabdf} 
\begin{tabular}{|c|c|c|} 
\hline 
$\beta$ & $D_f \,\,\,(L=64)$ & $D_f\,\,\,(L=128)$\\ 
\hline 
$0.0$ & $1.15\pm0.03$ & $1.14\pm0.02$\\ 
$0.005$ & $1.18\pm0.03$ & $1.16\pm0.02$\\ 
$0.05$ & $1.20\pm0.03$ & $1.17\pm0.02$\\ 
$0.5$ & $1.22\pm0.02$ & $1.20\pm0.02$\\ 
$5$ & $1.26\pm0.02$ & $1.24\pm0.02$\\ 
\hline 
\end{tabular} 
\end{centering} 
\end{table}  
\begin{table} 
\begin{centering} 
\protect \caption{Fractal dimension of the QBM model for different  
values of $\eta$ and $a$ for clusters of size  
t$L=64$. The values of the other parameters are:  
$\beta=0.5$ and $\alpha=1$.} 
\label{tabdf1} 
\begin{tabular}{|c|c|c|c|} 
\hline 
$\eta$ & $D_f(\eta,a=1)$ & $a$ & $D_f(\eta=1, a)$\\ 
\hline 
$0.5$ & $1.36\pm0.03$ & $0.5$ & $1.33\pm0.03$\\ 
$1.0$ & $1.21\pm0.03$ & $1.0$ & $1.21\pm0.03$\\ 
$2.0$ & $1.13\pm0.03$ & $2.0$ & $1.15\pm0.03$\\ 
$3.0$ & $1.10\pm0.03$ & $3.0$ & $1.10\pm0.03$\\ 
\hline 
\end{tabular} 
\end{centering} 
\end{table}  
\begin{table} 
\begin{centering} 
\protect \caption{Fractal dimension of the QBM model for different  
values of $\eta$ and $a$ for clusters of size  
$L=128$. The values of the other parameters are:  
$\beta=0.5$ and $\alpha=1$ } 
\label{tabdf2} 
\begin{tabular}{|c|c|c|c|} 
\hline 
$\eta$ & $D_f(\eta,a=1)$ & $a$ & $D_f(\eta=1, a)$\\ 
\hline 
$0.5$ & $1.42\pm0.02$ & $0.5$ & $1.40\pm0.02$\\ 
$1.0$ & $1.20\pm0.02$ & $1.0$ & $1.20\pm0.02$\\ 
$2.0$ & $1.15\pm0.02$ & $2.0$ & $1.16\pm0.02$\\ 
$3.0$ & $1.11\pm0.02$ & $3.0$ & $1.10\pm0.02$\\ 
\hline 
\end{tabular} 
\end{centering} 
\end{table}  
 
\begin{table} 
\begin{centering} 
\protect \caption{Behavior of the backbone fractal dimension ($D_B$) and  
($D_C$) of chemical distance fractal dimension, with $\beta=0.5,  
\alpha=1, \eta=1, a=1$, for clusters of size $L=64$ and $L=128$.} 
\label{tabdf3} 
\begin{tabular}{|c|c|c|} 
\hline 
$L$ & $D_B$ & $D_C$\\ 
\hline 
$64$ & $1.07\pm0.02$ & $1.00\pm0.02$\\ 
$128$ & $1.10\pm0.02$ & $1.00\pm0.02$\\ 
\hline 
\end{tabular} 
\end{centering} 
\end{table}  

\end{document}